\documentclass[twocolumn,showpacs,epsfig,preprintnumbers,amsmath,amssymb]{revtex4}

\usepackage{graphicx}
\usepackage{dcolumn}
\usepackage{bm}
\usepackage{amssymb}
\def \be{\begin{equation}}
\def \ee{\end{equation}}
\def \bea{\begin{eqnarray}}
\def \eea{\end{eqnarray}}
\def \nn{\nonumber}

\def \ben{\begin{enumerate}}
\def \een{\end{enumerate}}

\def \expect#1{\langle#1\rangle}
\begin{document}

\title{Propagation of interacting force chains in the continuum limit}
\author{Yael Roichman and Dov~Levine}
\affiliation{Department of Physics, Technion, Haifa 32000, Israel}

\author{Irad Yavneh}
\affiliation{Department of Computer Science, Technion, Haifa
32000, Israel}

\date{\today}

\begin{abstract}
We study the effect of mergers in the force chain model describing
the stress profile in static granular materials. Combining
numerical and analytical calculations we show that granular
materials do not generally behave in an elastic-like manner,
however they may under specific conditions, which are elaborated.
Non-elastic behavior resulting from the non-linearity of the full
force chain model is discussed.
\end{abstract}

\maketitle

A striking characteristic of stress transmission in granular
matter is the network of highly singular lines, termed {\em force
chains}, along which stress propagates
\cite{liu95,clement00,behringer,dantu}. This force chain network
reflects the specific packing of the system, which is unique for
each experiment. Although this leads to significant fluctuations
in stress profiles from experiment to experiment, the average
stress profiles calculated over an ensemble of similar experiments
seem well defined \cite{behringer}. In particular, the ensemble
average of the response to a small localized force,  the {\em
response function} \cite{footnote1}, is a bell-shaped curve with
scaling properties similar to that of an elastic response function
\cite{reydellet,clement00,behringer,coppersmith}. It is natural,
therefore, to ask whether the average stress in granular materials
behaves according to elasticity theory.  In this Communication we
address this question within the context of the Force Chain Model
\cite{bouchaud} (See \cite{goldenberg} for a discussion of a model
of masses linked with linear and nonlinear springs.).  We find
significant deviations from elasticity, except for the case of an
isotropic packing with nearly isotropic applied forces.

The recently proposed Force Chain Model (FCM) \cite{bouchaud}
transforms the singular behavior of stress in states of a
granular material into a continuum theory by averaging over an
ensemble of states. This is done by writing a Master
equation for the average density of force chains, allowing force
chains to propagate, split and merge \cite{bouchaud,socolar02}.
Previously, the response function of the FCM was calculated in three different fashions:
simulating force chain propagation in small quenched disordered
media \cite{bouchaud}; calculating the constitutive relation for
granular materials on large scales using a splitting-only variant
of the force chain model \cite{bouchaud}; and linearizing a
specific discretized version of the model around a homogeneous
solution \cite{socolar02}. The response functions calculated by
the first two methods agreed qualitatively with the experimentally
measured response \cite{behringer,clement00}, exhibiting a bell
shaped peak, while the third method gave a transition from
a single to a double peak at an intermediate length
scale \cite{socolar02}.

A priori, there are reasons to expect granular materials to behave
non-elastically: They cannot sustain tensile
stresses, they rearrange when external loads are changed,
and they have no equilibrium
stress-free state with respect to which to define a displacement field (See \cite{goldenberg} for a
different view on these issues.).  Despite these considerations, the central result of
Reference \cite{bouchaud} was that, in the absence of force chain mergers,
granular materials behave in a quasi-elastic manner on large length
scales.  In
this Communication, we argue that the effect of force chain mergers is to change this:
Generically, granular materials do not behave
elastically.  The apparent elastic-like behavior found in experiments
\cite{behringer,clement00} is restricted to specific packing
geometries (i.e. isotropic) and to specific configurations of
applied loads (i.e. near-isotropic).

Stress profiles in the FCM may be calculated in three ways, each
adapted to a different length scale: Monte-Carlo simulation on
small scales; numerical solution of the discretized model of Ref.
\cite{socolar02} on intermediate scales, and calculation of the
constitutive relation of the full FCM on large scales.  We will
discuss the latter two methods; the simulation results will be
presented in \cite{next}.   We will show that observed
deviations from elasticity \cite{behringer} may be understood in
the context of the FCM.

In the framework of the FCM, a force chain is
characterized by its intensity, $f$ (the pressure exerted
on each grain along the force chain), its direction, $\hat{n}$ ,
(which is determined with respect to the applied
force on the boundary), and its position, $\vec{r}$. There are four
events that involve the creation or annihilation a force chain
 \{$f, \hat{n}$\} at $\vec{r}$
\cite{socolar02}: it can split;
another force chain can split, creating it as one of its offspring; it can merge
with another and so be annihilated; or two other force chains can
merge, creating it. This yields a Master equation for
the force chain density, $P\equiv P(f,\hat{n},\vec{r})$
\cite{bouchaud,socolar02}:
 \bea \label{prob}
&& \hat{n}\cdot \vec{\nabla} P =- \frac{1}{\lambda}P\\
&&+ \frac{2}{\lambda}\int
P_1\psi_0\delta(f\hat{n}-(f_1\hat{n}_1+f_2\hat{n}_2))d{\underline{ f_1}}d{\underline{ f_2}}\nn\\
&&-Q P \int P_1 \varphi_0 \delta(f_2\hat{n}_2-(f_1\hat{n}_1+f\hat{n}))d{\underline{ f_1}}d{\underline{ f_2}}\nn \\
&&+ \frac{Q}{2}\int
 P_1P_2 \varphi_0 \delta(f\hat{n}-(f_1\hat{n}_1+f_2\hat{n}_2))d{\underline{ f_1}}d{\underline{ f_2}}
 .\nn\eea
Here $P_i=P(f_i,\hat{n_i},\vec{r})$, $Q$ is the force chain width
\cite{footnote2}, $\lambda$ is the splitting mean free path of a
force chain, and the functions $\psi_0$ and $\varphi_0$ are the
weights of a splitting/merging event (which depend, in principle,
on the directions $\hat{n}, \hat{n}_1$, and $\hat{n}_2$).
$d{\underline{ f_j}} \equiv df_j d\hat{n}_j$, and the delta
functions ensure force balance. Throughout this Communication we
will assume that when two force chains meet, they merge, i.e.
$\varphi_0=1$.

Following \cite{bouchaud}, we
define the force chain intensity density, $F(\hat{n},\vec{r})=
\int P(f,\hat{n},\vec{r}) f df$ and its angular moments:

 \bea
J_{\alpha}(\vec{r})&=&a\int n_{\alpha} F(\hat{n},\vec{r}) d\hat{n}\\
\sigma_{\alpha\beta}(\vec{r})&=&aD\int n_{\alpha} n_{\beta}
F(\hat{n},\vec{r}) d\hat{n}; \eea  $a$ is the grain size,
and D the dimension of the system.
$\sigma_{\alpha\beta}$ is the local stress tensor \cite{edwards}
and $\vec{J}(\vec{r})$ can be
thought of as the average force chain current.

In order to gain insight on the mesoscopic scale of stress
profiles in a granular material, we approximate Eq.(\ref{prob}),
following Ref. \cite{socolar02}. We employ the discrete ordinate
method proposed by Chandrasekhar \cite{chand} for solving the
Radiative Transfer Equation.   This approximates the integrals in
Eq.~(\ref{prob}) by sums, by discretizing the directions of force
chains as:
$P(f,\hat{n},\vec{r})=\sum_{i=1}^{6}P_i\delta(f~-~f^*)\delta(\hat{n}-\hat{n}_i)$
%
%
 where $\hat{n}_i~=~(\cos{\theta_i},\sin{\theta_i}), i=1,..,6,$ and
$\theta_{i+1}-\theta_{i}=\frac{\pi}{3}$ (with $\theta_7 \equiv \theta_1$).
The $P_i$'s are six
different functions representing the weights of the force chains
propagating in directions $\hat{n}_i$. Note that the choice
$\theta_{i+1}-\theta_{i}=\frac{\pi}{3}$ implies that all forces have
the same intensity, $f^*$, in order to satisfy force balance.
Substituting this into Eq. (\ref{prob}) results in six
coupled differential equations for the six force chain densities,
$P_i$. Rescaling $P_i\rightarrow\frac{1}{\lambda
Q}P_i$ and $\vec{r}\rightarrow \frac{\vec{r}}{\lambda}$
we arrive at the dimensionless equations \cite{socolar02}: \bea
\label{desc1} \hat{n}_i\cdot\nabla P_i&=&
-P_i+P_{i+1}+P_{i-1}\nn\\
&+& P_{i+1}P_{i-1}-P_{i+2}P_{i}-P_{i-2}P_i \eea Note that these
equations are written for isotropic homogeneous assemblies, since
the mean free path is assumed constant.

In Ref. \cite{socolar02} homogeneous solutions of the form
$
\{P_j\} = \{q,q^2,q,q^{-1},q^{-2},q^{-1}\}
$
 for any $q$ were considered; however there are others, for example:
$
\{P_j\} = \{q^{-1},1,q,q,1,q^{-1}\}
 $.
In \cite{socolar02}, Eq. (\ref{desc1}) was solved by linearization
around the first homogeneous solution, and, remarkably, a double
peaked response was shown to emerge at intermediate depths. It is
important to understand what physical conditions these solutions
correspond to. Let us begin by choosing $q=1$ in either of the
homogeneous solutions; this means that the force chain density,
and therefore the stress, is uniform throughout. This represents
the case of an isotropic array stressed hydrostatically, which is
the reference frame for response function experiments. In the case
that $q\neq1$ we have some predetermined relation between force
chain densities in different directions at the same position which
is unphysical for an isotropic material.

\begin{figure}[h]
\includegraphics[scale=0.25]{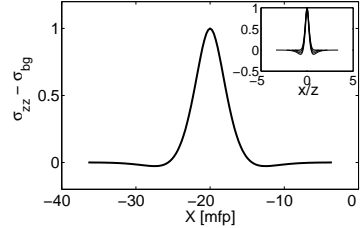}
\caption{The response function as calculated with boundary
conditions: $P_i=0$. The width of the response scales linearly
with depth as seen in the insert, where the response function at
various depths is plotted; the curves normalized by peak height.
See footnote \cite{footnote}. \label{numeric_resp}}
\end{figure}

\begin{figure}[h]
\includegraphics[scale=0.30]{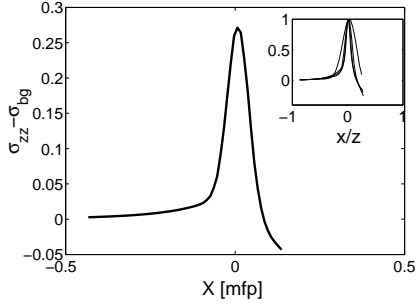}
\caption{\label{angle} The response of a granular assembly to a
force tilted $60^o$ with respect to the free surface. Insert:
Stress normalized by peak height, showing linear scaling with
depth. See footnote \cite{footnote}.}
\end{figure}

Numerical solutions of the descretized model, Eq. (\ref{desc1}),
were obtained using a second-order accurate finite-difference
approximation and solving the resulting nonlinear algebraic system
of equations iteratively \cite{next}. In Figure
\ref{numeric_resp}, the response to a normal force as calculated
by this model is plotted \cite{footnote}.  The peak is symmetric,
as expected.  In contrast,
the response to a tilted force (Figure \ref{angle} shows the
pressure profile on a plane normal to the applied force.) is
asymmetric. The width at half maximum of both functions increases
linearly, in accord with the experiments of Ref. \cite{behringer}.
The bell shaped peak of the response to the normal force is in
agreement with the elastic-like behavior ascribed to granular
materials (see {\em e.g.} \cite{savage,degennes,bouchaud}).
However, the asymmetric response to the tilted force deviates from
the elasticity prediction \cite{landau}.


An explanation for the deviation from elasticity of the response
function to a tilted force can be found in Eq. (\ref{prob})
which connects the force chain
density in one direction with that in any other
direction. This means that if all force chains arriving at the
surface are canceled (the grains rearrange to have zero stress at
a free surface, for instance) the total force chain density will
be zero in the vicinity of that surface. For an elastic material,
however, it is possible to have no strain
in the direction perpendicular to the surface and a finite strain
parallel to the surface, since the strain
components are independent. It is noteworthy that this deviation
from elasticity is observed also in the splitting-only version of
the force chain model (see \cite{guy}).

\begin{figure}[h]
\includegraphics[scale=0.35]{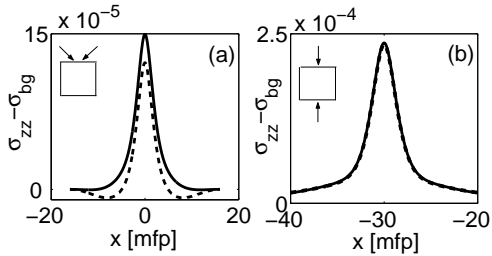}
\caption{\label{inta} A comparison between the response to two
forces applied simultaneously (dashed line), and applied
separately (solid line). a) the forces are applied close to each
other, b) the forces are applied at a distance. Lack of
superposition obtains for (a) but not for (b).  See footnote
\cite{footnote}.}
\end{figure}

One of the fundamental characteristics of the full force chain
model is its non-linear nature (see Equations (\ref{prob}) and
(\ref{desc1})).  In order to estimate the effect of this
non-linearity we tested superposition by comparing the response to
two different perturbations, first applied simultaneously and then
applied separately.  Figure \ref{inta}(a) presents the response to
two forces applied close to one another, and Figure \ref{inta}(b)
two forces applied further apart. It is clear that while the
effect of force chain interaction is significant in the former
case, it is negligible in the latter.

While these results are suggestive, the discretized model gives
incomplete understanding because all forces in the system are
equal. Thus, to better our understanding of the effect of force
chain mergers we calculated the constitutive relation on the
macroscopic scale by calculating angular moments of Eq.
(\ref{prob}), in the spirit of Ref. \cite{bouchaud}. Multiplying
Eq. (\ref{prob}) by $f\hat{n}$ and integrating over $f\hat{n}$ we
arrive at the force balance equation:
 \be \vec{\nabla}\cdot \sigma=\vec{F}_0
\nn\ee where $\vec{F}_0$ is an external body force.  In order to
calculate the second moment of Eq. (\ref{prob}), one has to
compute the integral:
$$
I \equiv \int
 P_1P_2 f \hat{n}_{\alpha}\hat{n}_{\beta}\varphi_0 \delta(f\hat{n}-(f_1\hat{n}_1+f_2\hat{n}_2))d{\underline{f}}d{\underline{ f_1}}d{\underline{ f_2}}
 $$
It has been shown both in experiments
\cite{nagel97,brockbank,nagel00} and in simulations
\cite{radjai96} that the probability distribution of forces,
$P(f)$,
has a maximum, and decays exponentially for larger forces.
Thus, we approximate the above integral 
by considering
small deviations of $f_1$ and $f_2$ from the intensity, $f_{max}$
at which the $P(f)$ is maximal. That is, we write:
\bea
f_1=f_{max}+\delta f_1\nn\\
f_2=f_{max}+\delta f_2 \eea and neglect terms with high orders
of $\delta f_i$ in 
$I$.  Moreover, we assume near-isotropy, by expanding
$F(\hat{n},\vec{r})$ in spherical harmonics and keeping only the
terms \be aF(\hat{n},\vec{r}) \simeq
p+D\hat{n}\cdot\vec{J}+\frac{D+2}{2}\hat{n}\cdot\tilde{\sigma}\cdot\hat{n}\\
\ee  where $\tilde{\sigma}$ is the traceless part of the stress tensor and
$p = \frac{1}{D} Tr \{\sigma\}$ is the pressure.   This
gives a constitutive relation
\cite{next}: \bea \label{stress}
\sigma_{\alpha\beta}&=&A\left[B(\phi)\vec{\nabla}\cdot
\vec{J}\delta_{\alpha\beta}+J_{\alpha\beta}\right]-C\expect{\hat{n}}\cdot
\vec{J}(\vec{r})\delta_{\alpha\beta}\nn\\
&-&D\left(\expect{\hat{n}}_{\alpha}J(\vec{r})_{\beta}+\expect{\hat{n}}_{\beta}J(\vec{r})_{\alpha}\right)\eea
where $\expect{n}=\equiv \int \hat{n}P(f,\hat{n},\vec r)
d{\underline{f}}$,
$J_{\alpha\beta}\equiv\frac{1}{2}\left(\partial_{\alpha}J_{\beta}+\partial_{\beta}J_{\alpha}\right)$,
and whose constants $A,C,D$ are determined by the specifics of the
granular packing.  The function $B(\phi)$ depends as well on the
force chain density $\phi(\vec r) \equiv \int P(f,\hat{n},\vec r)
d{\underline{f}}$ \cite{next}.

For nearly homogeneous and isotropic systems, the terms which are
products of $\vec{J}$ and $\expect{\hat{n}}$ are smaller than
terms linear in $\vec{J}$.  If they may be ignored, the
constitutive relation reduces to: \be
\sigma_{\alpha\beta}=A\left[B(\phi) \vec{\nabla}\cdot
\vec{J}\delta_{\alpha\beta}+J_{\alpha\beta}\right] \ee This
equation is formally equivalent to the constitutive relation of
conventional elasticity \cite{landau}. Therefore, we can define
two pseudo-elastic moduli: The pseudo-Poisson ratio,
$\nu=\frac{B}{1+2 B}$, and the pseudo-Young modulus, $E=A(1+\nu)$.
These depend not only on the geometry of the pile, but also on
position through the force chain density $\phi(\vec{r})$.
Therefore, the pseudo elastic behavior obtains only for nearly
homogenous systems.

Generally speaking, the constitutive relations calculated by the
force chain model (Eq. (\ref{stress})), are different from those
of conventional elasticity; in particular, they are nonlinear.
This nonlinearity is somewhat subtle, and holds for the ensemble
averaged stresses.  For a given packing, which does not change
upon application of external forces, it is clear that
superposition must hold, since the grain-scale equations of force
balance are linear.  Consider, however, the ensemble of stress
states which are consistent with given set of external forces
$\{{\cal F}_1\}$.  We believe that this ensemble is statistically
different from that ensemble which is compatible with a different
set of external forces $\{{\cal F}_2\}$, even if $\{{\cal F}_1\}$
and $\{{\cal F}_2\}$ are very similar. Physically, this is
reflected in the fragility \cite{cates99a} of the material:
rearrangements occur when the external conditions are changed.
Thus, the FCM predicts that if an ensemble averaged stress field
is measured, then granular materials will exhibit nonlinearity in
its response.


In this Communication we have dealt mainly with the ensemble
average of the stress profile in granular materials. However, the
singularity of the force chains and the wide distribution of force
chain intensity measured \cite{liu95} suggest that it might be
interesting to study force fluctuations, and the effect of
friction, in the framework of the FCM. As in References
\cite{bouchaud,socolar02}, the existence of force chains, in the
sense of a reasonably straight line of grains in contact, was
assumed.  It remains to be seen under what conditions this
assumption is reasonable.  We expect that for very hard grains
(more precisely, small stresses compared to the grain
compressibility), force chains will exist. If this is the case,
the effect of friction would be to change the details of the
packing obtained, such as the scattering mean free path and
persistence length, but not undermine the existence of force
chains (Indeed, grains are frictional in all real experiments
\cite{clement00,behringer,dantu}. Furthermore, friction might
stabilize a force chain network, by allowing it to bear loads
which would otherwise be "incompatible", this in turn can lead to
a larger load regime for which the granular material responds
linearly.

DL gratefully acknowledges support from the US-Israel Binational
Science Foundation (grant 9900235) and the Israel Science
Foundation (grant 88/02).  It is a pleasure to acknowledge useful
and interesting discussions with G. Bunin, P. Claudin, C.
Goldenfeld, I. Goldhirsch, S. Rahav, J. Socolar, and Y. Srebro.


\begin{thebibliography}{10}
\bibitem{footnote2} $Q$, which has dimensions of $length^{d-1}$ in $d$ dimensions, is analogous to
the chain cross section.
\bibitem{footnote} Note that there a background threshold stress $\sigma_{bg}$ to the
stress distributions of the discretized model, which corresponds
to the fixed point solution $P_j = 1$.  This is subtracted in the
figures.
\end{thebibliography}
\end{document}